# Optical control of non-equilibrium superconducting phase transition below $T_C$ in a cuprate

Claudio Giannetti[1], Giacomo Coslovich[2], Federico Cilento[1], Gabriele Ferrini[1], Hiroshi Eisaki[3,4✢], Nobuhisa Kaneko[4,✢✢], Martin Greven[3,4] and Fulvio Parmigiani[2,5*]

*[1]Department of Physics, Università Cattolica del Sacro Cuore, Brescia I-25121, Italy. [2]Department of Physics, Università degli Studi di Trieste, Trieste I-34127, Italy. [3]Department of Applied Physics, Stanford University, CA 94305, USA. [4]Stanford Synchrotron Radiation Laboratory, Stanford, CA 94305, USA. [5]Sincrotrone Trieste S.C.p.A., Basovizza I-34127, Italy. ✢Present address: Nanoelectronics Research Institute, National Institute of Advanced Industrial Science and Technology, Tsukuba, Ibaraki 305-8568, Japan. ✢✢Present address: National Metrology Institute of Japan, National Institute of Advanced Industrial Science and Technology, Tsukuba, Ibaraki 305-8568, Japan.*

*To whom correspondence should be addressed. E-mail: fulvio.parmigiani@elettra.trieste.it

**Photoinduced phase transitions from insulating to metallic states, accompanied by structural re-arrangements, have been recently reported in complex transition-metal oxides. However, the optical control of a purely electronic phase transition, where the thermodynamic phase is determined by the distribution of excitations, has remained elusive. Here we report optical control of the electronic phase in an underdoped $Bi_2Sr_2CaCu_2O_{8+\delta}$ crystal through impulsive photoinjection of quasiparticles (QP) via ultrashort laser pulses, avoiding significant laser heating. An abrupt transition of the transient optical electronic response is observed at a critical fluence of $I_{pump} \cong 70$ μJ/cm$^2$. We show that the measured dynamics is consistent with an inhomogenous first-order superconducting-to-normal state phase transition, triggered by a sudden shift of the chemical potential.**

The impulsive modification by ultrashort light pulses (~100 fs) of electronic, magnetic and structural properties in complex materials, has recently opened the field of photo-induced phase-transitions (PIPT)(1). Previous studies have focused on the insulator-to-metal PIPT in perovskite manganites (*2-4*) or vanadium dioxide (*5,6*), in which the transformation of both the electronic and structural properties occurs at once. On the contrary no evidence of optical control of a purely electronic phase transition has been reported so far.

The superconducting-to-normal state phase-transition (SNPT) is one of the most important examples of electronic phase transitions in solid-state systems. At zero temperature the BCS theory (*7*) explains the formation of the superconducting (SC) electronic phase in terms of the macroscopic condensation of the Cooper pairs, originated from the phonon-mediated coupling of two electrons. The excitation spectrum is characterized by the opening of the superconducting gap $\Delta_k$, representing half the energy necessary to break a Cooper pair and create two excitations with wavevectors +**k** and -**k**, i.e. quasiparticles (QP) (*7*). At finite temperature the main role of the electronic distribution is to block states otherwise available to form the superconducting condensate. As the temperature of the system is increased, the electronic distribution causes the continuous closing of the SC gap $\Delta_k(T)$ up to the point where $\Delta_k(T_C)=0$ and the second-order SNPT takes place.

The basic idea of non-equilibrium superconductivity is to photo-inject an excess of excitations through an ultrashort pump pulse, modifying the equilibrium electron distribution on a timescale faster than the impulsively-induced thermal heating. Under these conditions the free energy of the superconducting system $F_{SC}(T,n)$ can be varied along non-equilibrium pathways, by suddenly varying the number of excitations *n*, whereas for thermal phase transitions $n(T,\mu=E_F)$ is univocally determined by the temperature and the chemical potential, through the Fermi-Dirac statistics (*8*). Non-equilibrium superconductivity has been studied in BCS systems, evidencing the possibility of a non-thermal quenching of the superconducting phase, inducing a first-order phase transition to a metastable, inhomogenous state (*9,10*). However, these experiments have been performed without sufficient time resolution to follow the onset of the SNPT.

The unconventional SNPT in high-$T_C$ superconductors (HTSC) is a physical phenomenon that has been the subject of intense studies (*11*) yet remains poorly understood. Broadly, the fascinating physics of HTSC is characterized by four unexplained features: (i) the ground state of the undoped compounds (*12*); (ii) the pseudogap phase (*13*) where Cooper pairs are probably formed locally in the absence of a macroscopic condensate (*14*); (iii) the intrinsic inhomogeneity of the SC phase at the nanometer scale (*15*); (iv) the basic interaction responsible for the binding potential necessary to form Cooper pairs (*16*). So far, the SC phase has been studied either at equilibrium in the frequency domain (*17*) or out of equilibrium in the time-domain by externally modifying the temperature of the system (*18, 19*).

The control of the dynamics of the SNPT in general and, in particular, in high-$T_C$ cuprates, has never been achieved by direct interaction with photons, mainly due to problems related to sample heating (*19*). In the cuprates, the condensation energy of the SC state is small as compared to the energy gain of the insulating phase in complex metal oxides and one does not expect significant changes of the optical properties in the visible spectral range, where femtosecond resolutions are generally achieved. The delicate balance between the small variations of the optical properties and the low threshold expected (*18, 20*) to photoinduce a merely electronic SNPT has prevented scholars from investigating this physics.

In this work we perform time-resolved reflectivity measurements in the near-infrared with a tunable repetition rate femtosecond laser source. By setting the proper repetition rate (<100 kHz), we are able to impulsively excite an underdoped $Bi_2Sr_2CaCu_2O_{8+\delta}$ crystal ($T_C$=81 K) with pump fluences ranging from 10 μJ/cm$^2$ to 700 μJ/cm$^2$ while avoiding a significant heating of the material. Under these conditions, above a critical fluence of $I_{pump}\cong 70$ μJ/cm$^2$, we observe an abrupt transition of the electronic response. This behavior is an unambiguous signature of a photoinduced phase transition from the superconducting state to a new metastable state.

Fig. 1A shows the time-resolved reflectivity variation measured at room temperature. The impulsive variation, at zero delay, reflects the non-equilibrium excitation of the electronic system, due to the pump pulses. In agreement with the literature (*21*), the decay of the reflectivity in the normal state is well described by the two temperature model, in which the fast dynamics of ~350 fs is related to the electron-phonon relaxation. The inset of Fig. 1A displays the maximum reflectivity variation ΔR/R(0) as a function of the pump fluence. A linear dependence is observed over the entire experimentally available range. When the sample is cooled below $T_C$, the relaxation time increases up to few picoseconds, as shown in Fig. 1B. This well-established (*18, 22*) decrease of the decay rate of photo-injected excitations below $T_C$ is generally attributed to the suppression of the QP recombination process related to the opening of the superconducting gap (*23, 24*) (Fig. 2) and, on the μs timescale, to the energy and momentum conservation rules of nodal QP (*25*). These observations confirm that time-resolved reflectivity, in the visible spectral range, is sensitive to the density of the photoinjected excitations (*19, 20*). We also note that at $T<T_C$, the rise time of the ΔR/R signal increases up to ~300 fs. The inset of Fig. 1B reports the dependance of ΔR/R(0) on the pump intensity, at a repetition rate of 108 kHz. The fluence range so far explored (*19*) with higher repetition rate sources is highlighted by the red square. Outside this region, at repetition rates higher than 100 kHz, the average heating of the sample can not be prevented. In this region we measured a linear scaling of the signal. At larger fluences we observed a progressive saturation of the signal up to the point at which a different linear behaviour sets in. Exactly the same phenomenon is observed repeating the measurements at half the repetition rate (54 kHz), i.e. with the same fluence but with half the average power impinging on the sample. This observation excludes disturbing effects connected to the average heating of the excited area.

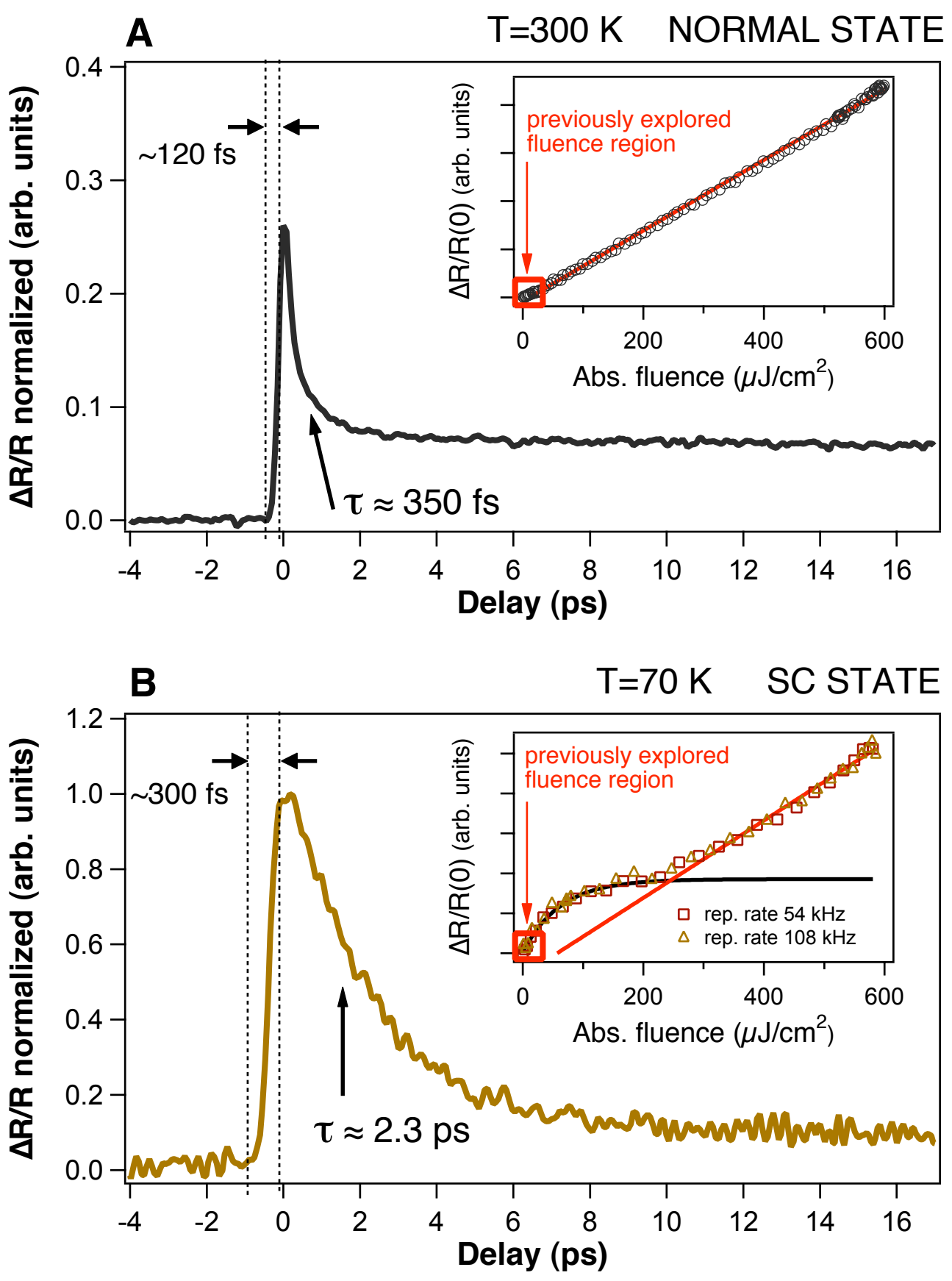

**Fig. 1** Time-resolved reflectivity variation on an underdoped $Bi_2Sr_2CaCu_2O_{8+\delta}$ crystal with $T_c$=81 K. (**A**) Measured relaxation dynamics at T=300 K (black line). The electron-phonon thermalization is completed within 350 fs after the pump excitation. The inset shows that the reflectivity variation at t=0 (ΔR/R(0)) scales linearly (red line) with the absorbed pump fluence in all the experimental available intensity range. (**B**) Measured relaxation dynamics below the critical temperature (yellow line). The measurement has been performed with $I_{pump}\cong 23$ μJ/cm², obtaining ΔR/R(0)≈$2.5\cdot 10^{-5}$. The recombination of the photo-injected excitations is quenched by the opening of superconducting gap and the recovery dynamics is completed on the picosecond time-scale. The inset shows that ΔR/R(0) is linear with the absorbed fluence in the range previously explored in the literature through high-repetition rate sources. Extending the fluence range we observe a saturation of the ΔR/R(0) signal, followed by a linear behaviour at high intensities. The black and red lines are the fit to the data, obtained through a saturation and a linear function, respectively.

To interpret this result we report in Fig. 3A the femtosecond detail of the time-resolved reflectivity in the 10-300 μJ/cm² fluence region. At low fluences the data can be fitted (gray lines superimposed on the data) with the numeric result of the integration of the Rothwarf-Taylor equations (RTE) (*26*), describing the number *n* of excitations coupled to non-equilibrium gap-energy phonons (Supporting Online Material), in agreement with the results reported in the literature (*18, 23, 24*). In this intensity regime, the results of the fitting procedure indicate that the pump energy is mainly absorbed through excitation of the phonon population. This finding satisfactorily explains both that the decay dynamics is dominated by the inelastic scattering rate of high-energy phonons ($\gamma_{esc}\approx 2$ ps$^{-1}$ (*21*)) without any intensity dependence (*19*) and that a constant delay of ~300 fs of the maximum of the signal is measured (*23*). At absorbed energy densities larger than ~70 μJ/cm², a fast peak emerges (Fig. 3B) and causes a decrease of the signal rise-time (Fig. 3C), while the following slow dynamics results progressively delayed (Fig. 3D). Above this intensity threshold the data can no longer be fitted with the RTE.

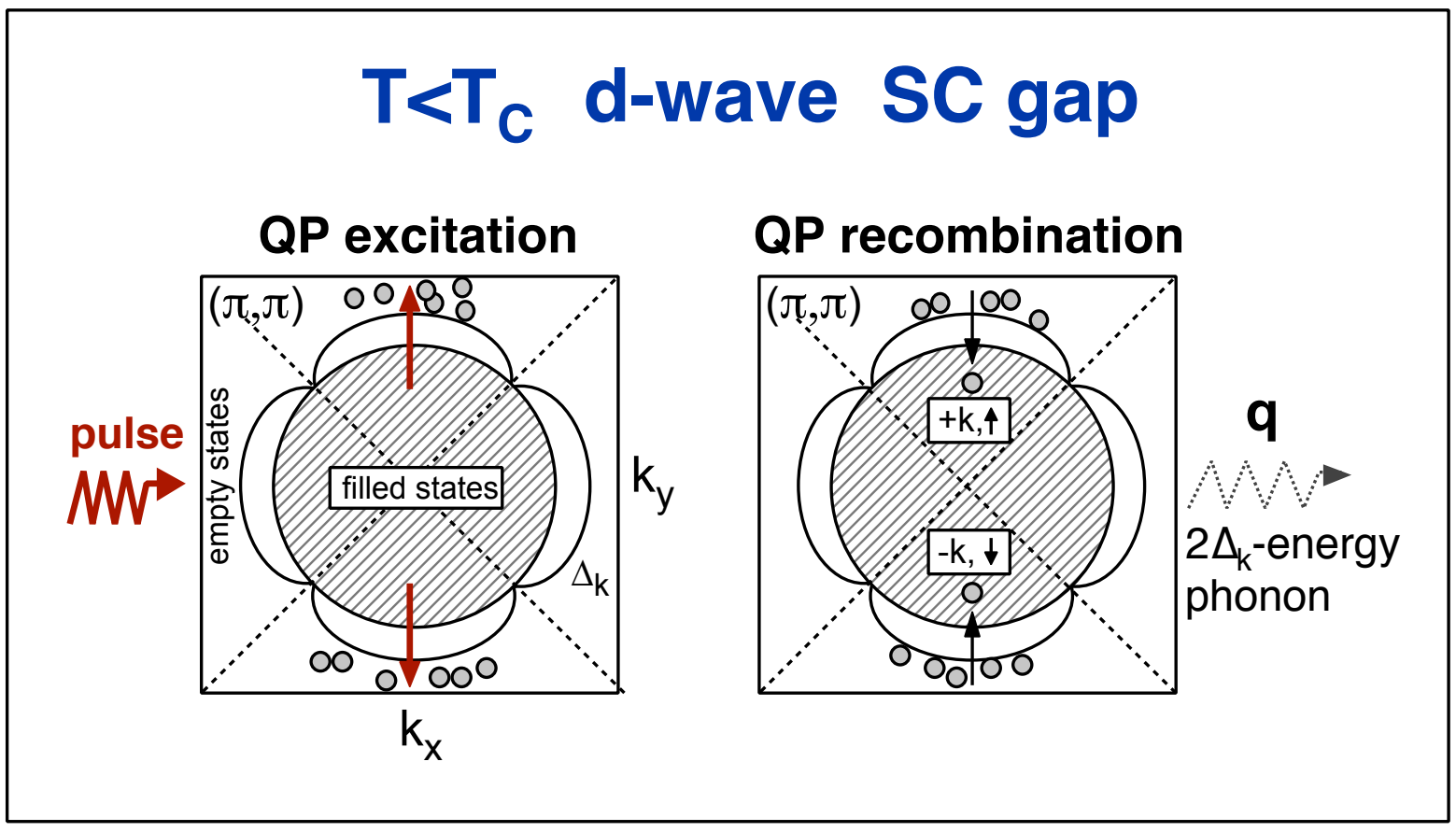

**Fig. 2** Rough scheme of the excitation and relaxation processes of antinodal QP in the **k**-space for a two-dimensional BCS-like superconductor with a spherical Fermi surface and a d-wave gap symmetry. The superconducting phase at $T<T_C$ is characterized by a gap $\Delta_k$ in the excitation spectrum. The four nodes in the (π,π) directions of the **k**-space are the consequence of the d-wave symmetry of the gap. Let us consider the decay process via scattering between two quasi-particles, with energy $E_{k1}$, $E_{k2}$ and momentum $\mathbf{k_1}$, $\mathbf{k_2}$, and a phonon with momentum **q** and energy $\hbar v_s|\mathbf{q}|$, $v_s$ being the sound velocity. The momentum and energy conservations imply: $\mathbf{k_1}+\mathbf{k_2}=\mathbf{q}$ and $E_{k1}+E_{k2}=\hbar v_s|\mathbf{q}|$. The constraints related to the simultaneous energy and momentum conservation quench both the decay of QP towards the nodes, via emission of two additional QP (Gedik2004), and the nodal QP recombination through emission of a gap-energy phonon (Feenstra1997). As a consequence, the photo-excitation leads to the accumulation of QP on the top of the SC gap at the antinodes, within ~1 ps. When two nodal QP recombine, a Cooper pair with momentum **k** is formed and a $2\Delta_k$-energy phonon is emitted.

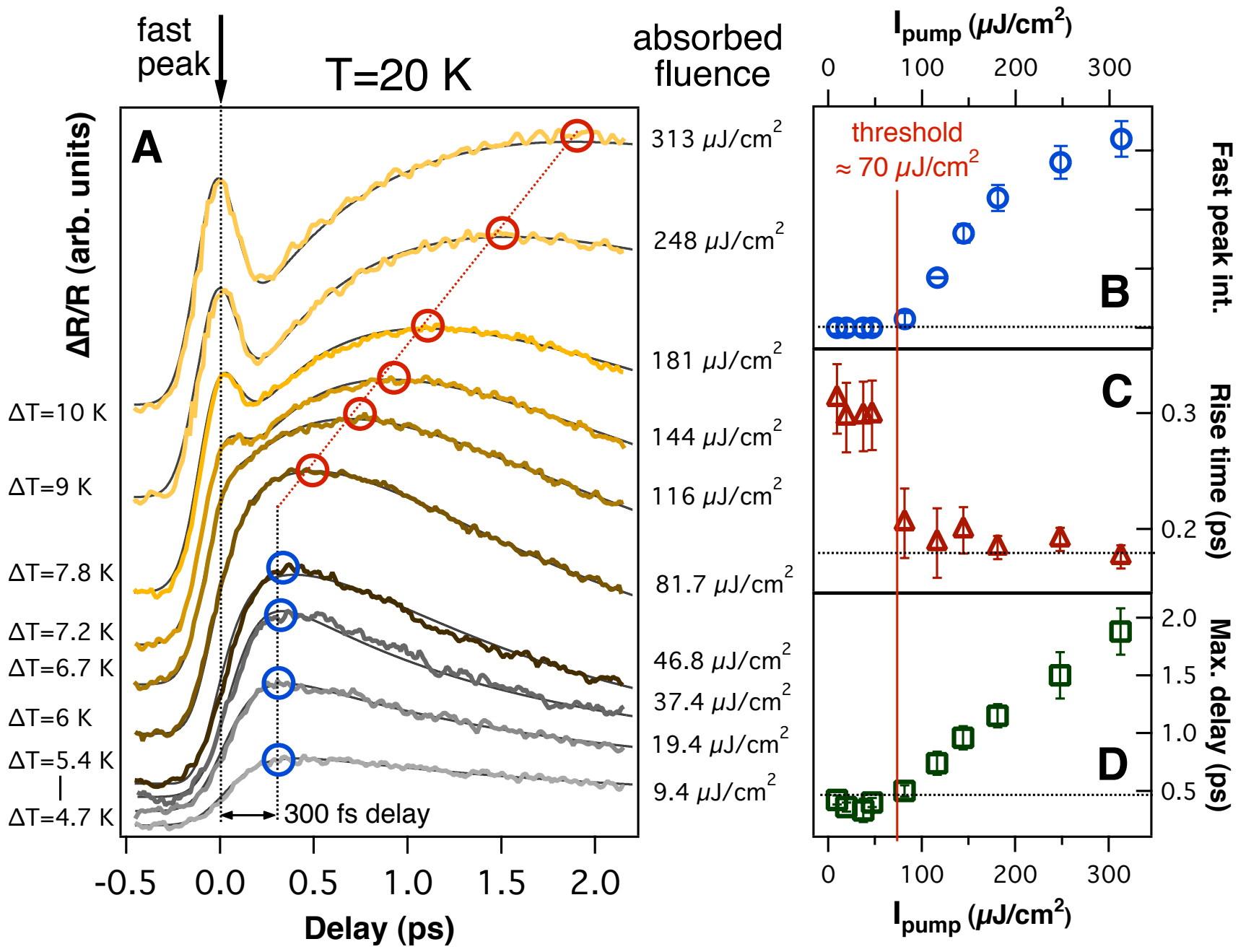

**Fig. 3** Time-resolved reflectivity measurements at T=20 K and 108 kHz repetition rate. **(A)** An abrupt transition of the relaxation dynamics is measured at $I_{pump}$>70 μJ/cm². Below this threshold the data are fitted with RTE. Above the threshold, a fast peak appears and the maximum of ΔR/R signal moves to positive delays. On the left, the maximum estimated average temperature increase due to both pump and probe beams is reported. The product of the RTE solution with the coalescence dynamics (Supporting Online Material) is fitted to the data (gray lines) on the whole experimental range. **(B)** Intensity of the fast peak, after subtraction of the slow dynamics, as a function of the pump fluence. **(C)** Decreasing of the rise time of the signal across the photo-induced SNPT. **(D)** Position of the maximum of the ΔR/R signal, as the pump fluence is increased.

The numerical solution of the heat equation and the calculated temperature increase profile, show that, at the threshold fluence, the maximum temperature increase is <10 K. In Fig. 4 we report the time-resolved reflectivity data on the same sample, at T>70 K. When the laser repetition rate is increased from 54 kHz to 540 kHz and the estimated average temperature exceeds $T_c$=81 K, the slow dynamics disappears and a flat signal after the fast peak, is measured. This is regarded as the evidence that the slow increase of ΔR/R, at $I_{pump}$>70 μJ/cm² and shown in Fig. 3A, originates from a local average temperature smaller than $T_C$. Therefore, the observed transition of ΔR/R at $I_{pump}$>70 μJ/cm² is a genuine signature of a non-thermal phase transition and the slow increase of the signal, on the ps timescale, is related to the recovery of the superconducting phase.

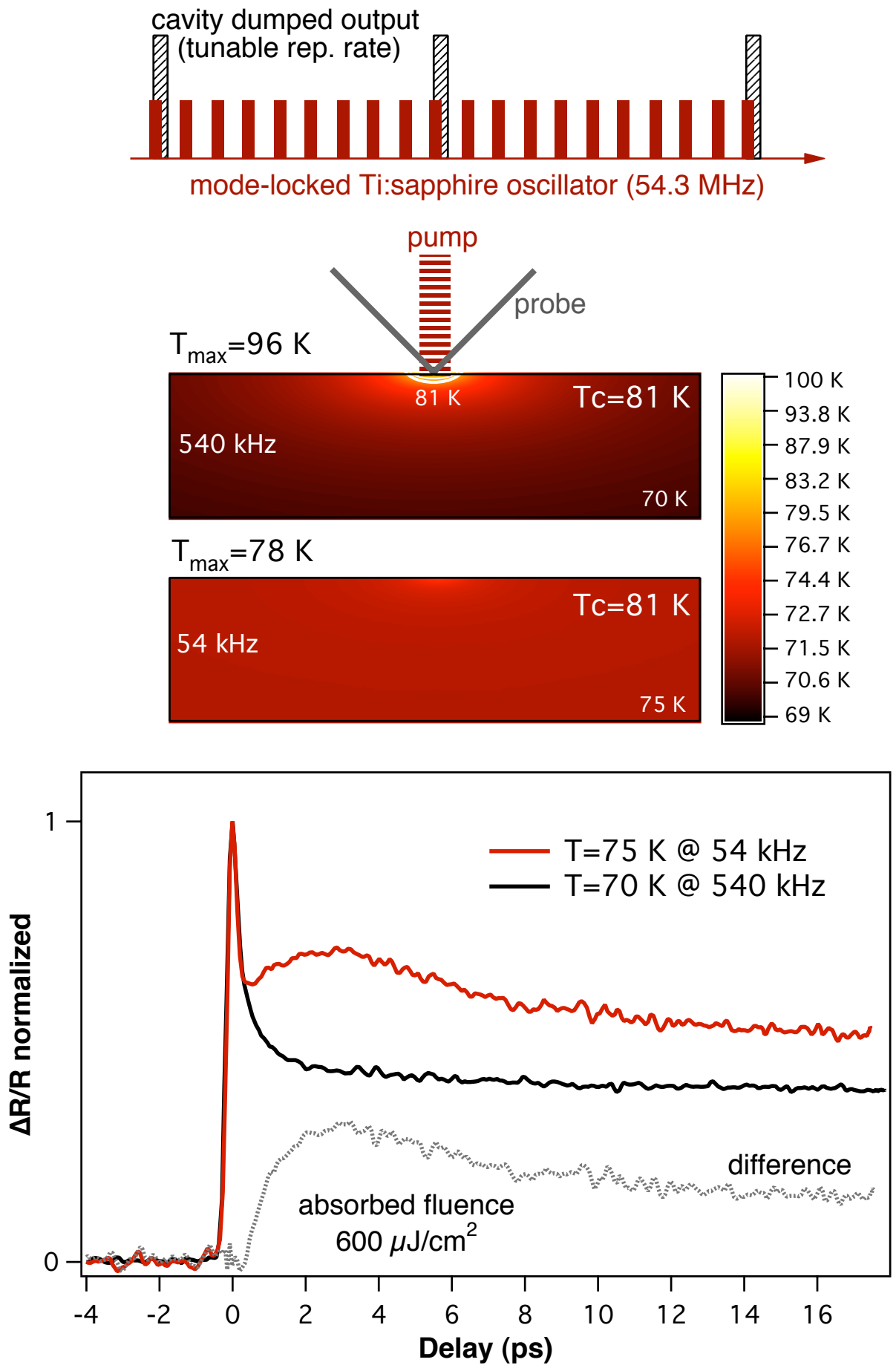

**Fig. 4** Effects of the thermal second order SNPT, induced by average heating, on the time-resolved measurements. The spatial profile of the temperature increase is calculated solving the heat equation on a 1 mm slab in thermal contact with the cryostat cold finger, at constant temperature T, and irradiated by both pump and probe beams. At 540 kHz repetition rate and $I_{pump}$=600 μJ/cm², the temperature of the volume probed exceeds $T_C$=81K, whereas at 54 kHz the average temperature is limited to 78 K. The graph displays the time-resolved reflectivity measurements at different repetition rates. When the repetition rate is increased and the average temperature exceeds $T_C$, the slow dynamics after the fast peak disappears and the decay dynamics becomes similar to the normal state one, reported in Fig. 1A. The dotted line shows the difference between the two normalized transient reflectivities.

Since ΔR/R is related to the excitations number (*18*), we conclude that, after the photo-induced phase transition, the number of excitations during the superconductivity recovery increases, reaching a maximum and decreasing again on the picosecond timescale (Fig. 3A). This is in contrast with the kinetics of the order parameter η in a second-order phase transition (*27*). When the average number of Cooper pairs, related to the order parameter, is impulsively decreased, at $T<T_c$ the system monotonically recovers the equilibrium value $\underline{\eta}$. This reflects the fact that the reconstruction of the SC gap implies a monotonic decrease of the excitations in the system. The measured increase of the signal after the fast decay suggests that an inhomogeneous metastable state is formed with the SC fraction increasing in time.

The perturbation of the electron distribution is generally mimicked either by introducing an effective temperature $T_{eff}$ (*28*) or an effective chemical potential $\mu_{eff}$ (*29*). Both the models predict a decrease of the superconducting gap $\Delta(n,T)$ with the excitation number, but with important differences. In the $T_{eff}$ model the gap dependence can be approximated by $\Delta(n,0)/\Delta(0,0)\approx 1-32(3n)^{3/2}/\pi^3$, $n$ being measured in units of $4N(0)\ \Delta(0,0)$ (*8*). A complete closing of the gap is obtained at $n_{cr}\approx 0.33$, causing a second-order phase transition to the normal state (*8*). In the $\mu_{eff}$ model the gap closing is slower and, before the complete collapse at $n\approx 0.65$, the free energy of the superconducting state equals the normal state one, causing a first-order phase transition from the superconducting to the normal state (*8*). We underline that the differences between the predictions of the two models are not related to the number of excitations injected into the system, but to their energy distribution. We compared our results with the predictions of the $T_{eff}$ model, performing the numeric integration of the gap equation in the d-wave symmetry and calculating the absorbed energy, taking into account the energy spent to create the QP, the variation of the energy of the superconducting condensate due to QP and the energy absorbed by the phonon system (*8*). The total absorbed intensity necessary for the complete closing of the SC gap is at least 7 times larger than the measured intensity threshold, allowing us to further exclude, on a quantitative base, a second-order phase transition as the origin of the experimental evidences. At variance with the $T_{eff}$ model, the inclusion of the phonon contributions in the $\mu_{eff}$ model is not straightforward, because the variation of the chemical potential describing the non-equilibrium distribution function does not affect the Bose-Einstein distribution of the lattice excitations. However, within the $\mu_{eff}$ model, the critical excitation number predicted is $n_{cr}\approx 0.16$, corresponding to an energy threshold smaller than the measured one. The description of the non-equilibrium excitation distribution through $T_{eff}$ fails because the energy distribution of the photoinjected excitations is flat during the action of the pump pulse and because the QP thermalization and recombination processes are suppressed by energy-momentum conservation and by the presence of the superconducting gap (*24, 25*). The combination of these two mechanisms leads to the accumulation of antinodal quasiparticles on the top of the gap. This non-thermal energy distribution causes a first-order phase transition to the normal state before the complete closing of the gap predicted by the $T_{eff}$ model can take place (*8, 29*).

Within this framework, the slow increase of the transient reflectivity, after the fast dynamics, contains the information on the final state after the impulsive photo-excitation and on the superconducting phase recovery. The observed dynamics can be interpreted as the signature of a coalescence process triggered by the photo-induced first-order phase transition. Within the first hundreds of femtoseconds the free energy of the superconducting phase equals the free energy of the normal phase, leaving the system in an inhomogenous mixture of the two phases. In the superconducting volume ($V_{SC}$) the relaxation dynamics is slow and described by the RTE, whereas in the normal fraction ($1-V_{SC}$) the decay of the excitations is extremely fast and is completed in a few hundreds of femtoseconds. On the picosecond timescale the metastable system can thus be schematized as superconducting regions surrounded by a normal volume where the free-energy of the superconducting phase is smaller than the free energy of the normal state and a coalescence process can start, in analogy with the case of a supersaturated solution (*27*). In this picture, the transient reflectivity is $\Delta R(t)/R\propto V_{SC}(t)\cdot n(t)+(1-V_{SC}(t))\cdot f_N(t)$, $f_N(t)$ being the transient reflectivity signal related to the normal phase (Fig. 1A). The dynamics of $V_{SC}(t)$ is modelled as the precipitation of the solute in a supersaturated solution (Supporting Online Material). At the beginning of the coalescence process the $\Delta R(t)/R$ signal is dominated by the growth of the superconducting volume, resulting in the increase of the signal, whereas on longer timescales the supersaturation tends to zero and the dynamics is regulated by the RTE. The fit of this model to the data is reported in Fig. 3A. A very good agreement with the measurements is obtained, confirming our interpretation.

The possibility to photoinduce a first-order electronic phase transition in a superconducting high-Tc system opens intriguing possibilities. First, the present measurements underline the interesting question of the real nature of the metastable final state after photo-excitation of the SNPT and its relationship with the underlying pseudogap state where the $\Delta R/R$ dynamics drastically changes (*22*). The detailed investigation of the electronic dynamics of the final state should help in discriminating between a pseudogap phase competing and co-existing with the superconducting phase. Second, a weak first-order SNTP is expected at $|T-T_C|/T_C \approx 10^{-6}$ K, because of the effects of the order parameter fluctuations (*30*). In the present experiment the first order phase transition is photoinduced by impulsively varying the order parameter, the thermodynamic temperature $T \ll T_C$ being fixed. This possibility opens the way to the study of new dynamical phases of non-equilibrium condensed-matter systems. Third, the rough coalescence model satisfactorily fits the data even at very high fluences where $t_0>0$, i.e. the SC phase is completely destroyed and the coalescence begins at positive delays. The lack of a signature of an earlier nucleation process could suggest a role played by the intrinsic inhomogeneity of the SC ground state at the nanometric scale (*15*). Fourth, a further investigation of the velocity of the first-order photo-induced SNPT can have important consequences on the understanding of the binding mechanism (*16*) (if there is any...) to form Cooper pairs in high-Tc. The presence of a bottleneck in the timescale of the SNPT could be the fingerprint of the coupling mechanism.

## References

1. K. Nasu, Ed., *Photoinduced Phase Transition* (World Scientific, Hackensack, NJ, 2004).
2. K. Miyano, T. Tanaka, Y. Tomioka, Y. Tokura, *Phys. Rev. Lett.* **78**, 4257–4260 (1997).
3. M. Rini *et al.*, *Nature* **449**, 72–74 (2007).
4. M. Matsubara *et al., Phys. Rev. Lett.* **99**, 207401 (2007).
5. A. Cavalleri *et al.*, *Phys. Rev. Lett.* **87**, 234701 (2001).
6. P. Baum, D.S. Yang, A. H. Zewail, *Science* **318**, 788-792 (2007).
7. P. Phillips, *Advanced Solid State Physics* (Westview Press, Boulder, CO, 2003).
8. E. J. Nicol, J. P. Carbotte, *Phys. Rev. B* **67**, 214506 (2003).
9. L. R. Testardi, *Phys. Rev. B* **4**, 2189-2196 (1971).
10. R. Sobolewski, D.P. Butler, T. Y. Hsiang, C.V. Stancampiano, G. A. Mourou, *Phys. Rev. B* **33**, 4604-4614 (1986).
11. W. S. Lee, Z. X. Shen, *Nature Phys.* **4**, 95-96 (2008).
12. S. Chakravarty, *Science* **319**, 735-736 (2008).
13. T. Timusk, B. Statt, *Rep. Prog. Phys.* **62**, 61-122 (1999).
14. S. R. Julian, M. R. Norman, *Nature* **447**, 537-538 (2007).
15. K. McElroy *et al.*, *Nature* **422**, 592-596 (2003).
16. P. Monthoux, D. Pines, G. G. Lonzarich, *Nature* **450**, 1177–1183 (2007).
17. D. N. Basov, T. Timusk, *Reviews of Modern Physics* **77**, 721-779 (2005).
18. V. V. Kabanov, J. Demsar, B. Podobnik, D. Mihailovic, *Phys. Rev. B* **59**, 1497-1506 (1999).
19. N. Gedik *et al., Phys. Rev. Lett.* **95,** 117005 (2005).
20. J. Demsar, R. D. Averitt, V. V. Kabanov, D. Mihailovic, *Phys. Rev. Lett.* **91**, 169701 (2003).
21. L. Perfetti *et al.*, *Phys. Rev. Lett.* **99,** 197001 (2007).

22. Y. H. Li *et al.*, http://arxiv.org/abs/0706.4282 (2008).

23. V. V. Kabanov, J. Demsar, D. Mihailovic, *Phys. Rev. Lett.* **95**, 147002 (2005).

24. N. Gedik, *et al. Phys. Rev. B* **70**, 014504 (2004).

25. B. J. Feenstra, J. Scüzmann, D. van der Marel, R. Pérez Pianya, M. Decroux, *Phys. Rev. Lett.* **79**, 4890-4893 (1997).

26. A. Rothwarf, B. N. Taylor, *Phys. Rev. Lett.* **19,** 27-30 (1967).

27. E. M. Lifshitz, L. P. Pitaevskii, *Physical Kinetics* (Butterworth-Heinemann Ltd, Oxford, 1981).

28. W. H. Parker, *Phys. Rev. B* **12**, 3667-3672 (1975).

29. C. S. Owen, D. J. Scalapino, *Phys. Rev. Lett.* **28,** 1559-1561 (1972).

30. B. I. Halperin, T. C. Lubensky, *Phys. Rev. Lett.* **32**, 292-295 (1974).

31. We thank F. Banfi for the fruitful discussions and for the help in simulating the average heating of the sample, S. Pagliara for assistance in the cryogenic set-up and A. Damascelli and Z.X. Shen for discussions. This work was supported by the Italian Ministero dell'Istruzione, Università e Ricerca (MIUR). The crystal growth work at Stanford University was supported by grants from the Department of Energy and the National Science Foundation.

# Supporting Material

### Experimental set-up

Time-resolved reflectivity measurements have been performed through a standard pump-probe set-up. The laser source is a cavity-dumped Ti:sapphire oscillator producing 120 fs-1.5 eV light pulses. The output energy is ~50 nJ/pulse at repetition rates tunable from 543 kHz to single shot. The sample is an underdoped $Bi_2Sr_2CaCu_2O_{8+\delta}$ crystal with $T_c$=81 K, characterized through AC magnetic susceptibility. The sample is mounted on a cold finger and covered by a coverslip to protect the surface during the cooling.

### Fitting procedures

The low fluence data have been fitted with the Rothwarf-Taylor equations (RTE),

$$\frac{dn}{dt} = I_n(t) + 2p\gamma - \beta n^2$$
$$\frac{dp}{dt} = I_p(t) - p\gamma + \frac{\beta}{2}n^2 - (p - p_{eq})\gamma_{esc} \quad (1)$$

describing the number of excitations $n$ coupled to phonons, $p$ being the gap-energy phonon number. The coupling of the electronic and phonon populations is obtained through a) the annihilation of a Cooper pair via gap phonon absorption ($p\gamma$ term) and b) the emission of gap phonons during the two-body direct recombination of excitations to form a Cooper pair ($\beta n^2$ term). The most relevant prediction of the model is that the decay dynamics of impulsive photo-injected electronic and phonon excitations ($I_n(t)$ and $I_p(t)$ source terms) is ultimately regulated by the escape rate of the non-equilibrium gap-phonons ($(p\text{-}p_{eq})\gamma_{esc}$ term). We fixed $\beta$≈0.1 $cm^2/s$ (*1*) and $\gamma$=3 $ps^{-1}$, as obtained from room temperature measurements.

### Average thermal heating

The average thermal heating has been simulated by integrating the heat equation on a 1 mm-thick slab, whose bottom temperature is kept equal to the cold finger one. Thermal insulation conditions are used on the other sides of the slab. The stationary solution has been calculated in cylindrical coordinates with a heat source modelled by:

$$I(x,y,z) = I_0 \alpha e^{-\alpha z} \frac{4\ln 2}{\pi fwhm^2} e^{-4\ln 2 \frac{r^2}{fwhm^2}} \quad (2)$$

$\alpha=6\cdot 10^6$ $m^{-1}$ being the absorption coefficient, *fwhm* the full-width-half-maximum of the beam profile, and $I_0$ the average power impinging on the sample. Both the pump and probe contributions and the temperature dependence of the anisotropic thermal conductivity (*2*) were taken into account.

**Coalescence**

The dynamics of the superconducting fraction $V_{SC}(t)$ is modelled by:

$$V_{SC}(t) = 1 - \left(\frac{c}{t - t_0 + c}\right)^{1/3} \quad (3)$$

where c and $t_0$ are coefficients related to the physical properties of the system and to the distribution of the grain sizes when the coalescence process starts at t=0 (*3*).

**References**

1. N. Gedik *et al., Phys. Rev. Lett.* **95,** 117005 (2005).
2. M. F. Crommie, A. Zettl, *Phys. Rev. B* **43**, 408 (1991).
3. E. M. Lifshitz, L. P. Pitaevskii, *Physical Kinetics* (Butterworth-Heinemann Ltd, Oxford, 1981).